\documentclass[twoside,onecolumn]{amsart}
\usepackage[T1]{fontenc}

\usepackage[english]{babel} 

\usepackage[hmarginratio=1:1,top=28mm,bottom=28mm,left=21mm,right=21mm,footskip=10mm]{geometry} 
\usepackage[hang, small,labelfont=bf,up,textfont=it,up]{caption} 
\usepackage{booktabs} 

\usepackage{abstract} 

\usepackage{enumerate}

\usepackage[]{titlesec}
\titleformat{\section}[block]
  {\fontsize{11}{10}\bfseries\sffamily}
  {\thesection}
  {1em}
  {}
\titleformat{\subsection}[block]
  {\fontsize{11}{10}\bfseries\sffamily}
  {\thesubsection} 
  {1em}
  {}
  
\usepackage{fancyhdr} 
\usepackage{titling} 
\usepackage[noblocks]{authblk}

\usepackage{hyperref} 
\usepackage{natbib}

\usepackage{url,lineno,microtype,subcaption}
\usepackage[onehalfspacing]{setspace}

\usepackage{inputenc,crop,graphicx,amsmath,array,color}
\usepackage{amssymb,flushend,stfloats,amsthm,chngpage,parskip,epstopdf}
\usepackage{float}
\usepackage[htt]{hyphenat}

\pretitle{\begin{center}\vspace{10pt}\LARGE\bfseries} 
\posttitle{\vspace{16pt}\end{center}} 

\title{Long-term monitoring of the broad-line region properties in a selected sample of AGN} 
\date{}
\author[1,*]{Dragana Ili\'c}
\author[2]{Alla I. Shapovalova}
\author[1,3]{Luka {\v C}.~Popovi\' c}
\author[4]{Vahram Chavushyan}
\author[2]{Alexander N. Burenkov}
\author[5]{Wolfram Kollatschny}
\author[1]{Andjelka Kova\v cevi\'c}
\author[1,3]{Sladjana Mar\v ceta-Mandi\'c}
\author[1,6]{Nemanja Raki\'c}
\author[7]{Giovanni La Mura}
\author[7]{Piero Rafanelli}

\begin{tiny}
\affil[1]{Department of Astronomy, Faculty of Mathematics, University of Belgrade, Studentski trg 16, 11000 Belgrade, Serbia}
\affil[2]{Astrophysical Observatory of the Russian Academy of Science, Nizhnij Arkhyz, Karachaevo-Cherkesia 369167, Russia}
\affil[3]{Astronomical Observatory, Volgina 7, 11060 Belgrade 74, Serbia}
\affil[4]{Instituto Nacional de Astrof\'{i}sica, \'{O}ptica y 
Electr\'{o}nica, Apartado Postal 51, CP 72000, Puebla, Pue. M\'{e}xico}
\affil[5]{Institut fuer Astrophysik, Universitaet Goettingen, Friedrich-Hund Platz 1, 37077, G\"{o}ttingen, Germany}
\affil[6]{Faculty of Science, University of Banjaluka, Mladena Stojanovi\'ca 2,78000 Banjaluka, Republic of Srpska, Bosnia and Herzegovina}
\affil[7]{Department of Physics and Astronomy, University of Padova, Vicolo dell'Osservatorio 3, I-35122 Padova, Italy}
\affil[*]{dilic@matf.bg.ac.rs} 

\end{tiny}

\begin{document}
\maketitle\date{\vspace{-10ex}}

\vspace{-2ex}
\begin{abstract}
We present the results of the long-term optical monitoring campaign of
active galactic nuclei (AGN) coordinated by the Special Astrophysical 
Observatory of the Russian Academy of Science. This campaign has produced a remarkable
set of optical spectra, since we have monitored for
several decades different types of broad-line (type 1) AGN, from a Seyfert 1, double-peaked line,
radio loud and radio quiet AGN, to a supermassive binary black hole candidate. 
Our analysis of the properties of the broad line region (BLR) of these objects is based 
on the variability of the broad emission lines. We hereby give a comparative review of the variability 
properties of the broad emission lines and the BLR of seven different type 1 AGNs, emphasizing some important results, such as the variability rate, the BLR geometry, 
and the presence of the intrinsic Baldwin effect. 
We are discussing the difference and similarity in the continuum and emission line variability,
focusing on what is the impact of our results to the supermassive black hole mass 
determination from the BLR properties. 

\vspace{6pt}\small{\bf keywords}: galaxies:active-galaxies, quasar:individual (Arp 102B, 3C 
390.3, NGC 5548, NGC 4151, NGC 7469, Ark 564, E1821+643), quasars: supermassive black holes, quasar:emission lines, line:profiles
\end{abstract}



\section{Introduction}

The broad line region (BLR) in active galactic nuclei (AGN) is responsible for the emission of intensive broad lines, seen in the optical spectra of type 1 AGN. We use these lines to probe the physics and kinematics of the BLR, aiming to study in more details the central engine and the supermassive black hole (SMBH) properties. This is because, on one hand the BLR is photoionized by the central continuum source, and on the other, it is gravitationally bound to the SMBH \citep{OF06}. 

One way to do this is by using the variability of the broad emission lines, as they are known to be strongly variable in flux and profile. From the long-term monitoring campaign of optical broad emission lines in different type 1 AGN, it is possible to estimate the size of the BLR \citep[reverberation analysis][]{LC72, BM82, GS86} and the mass of the SMBH \citep{Ga88}, and test the photoionization hypothesis by studying the line and continuum flux correlations \citep{OF06}. One particularly interesting correlation, first detected in the broad C IV line, is the Baldwin effect, i.e. the anti-correlation between the emission line equivalent width and the near-by continuum luminosity \citep{Ba77}. { There are two different types of Baldwin effect: i) global Baldwin effect which shows the above anti-correlation on single-epoch observations of a large number of AGN; and ii) intrinsic Baldwin effect which is the same anti-correlation detected in an individual variable AGN.} It is well known that the global Baldwin effect is not detected in case of the broad H$\beta$ line. However, when testing this correlation in case of a single object monitoring data, i.e. the intrinsic Baldwin effect, it has been seen that the anti-correlation exists  even in the case of H$\beta$ \citep{PP92, Ra17}.

Here we present some interesting findings of the long-term (more than a decade) optical monitoring campaign of a sample of seven type 1 AGN of different spectral types: Seyfert 1 galaxies (NGC 5548, NGC 4151, NGC 7469), Narrow-line Seyfert 1 galaxy - NLSy 1 (Ark 564), double-peaked line radio loud (3C 390.3) and radio quiet (Arp 102B) galaxy, and a luminous quasar (E1821+643), a supermassive binary black hole candidate. All spectral data for each objects have been first presented and analyzed separately in \cite{Sh01, Sh04, Sh08, Sh10a, Sh12, Sh13, Sh16, Sh17}. However, since the obtained spectra are observed, reduced and analyzed in the same manner, the final result is a uniform set of data, which can be compared and discussed.
Here we continue the discussion given in \cite{Il15}, where we first tried to give a comparative review of the variability properties of type 1 AGN in our sample, only that now our sample consist of seven objects, including one high-luminosity local quasar, which makes our sample more diverse. Here we focus on discussing the possible BLR geometry and the presence of the intrinsic Baldwin effect. The paper is organized as follows: in Section 2 we briefly  describe the observations, the obtained data and performed analysis; in Section 3 we present the results, and in Section 4 we give the discussion and outline our conclusions.

\section{Observations and data analysis}

The long-term monitoring campaign was coordinated by the Special Astrophysical Observatory of the Russian Academy of Science (SAO), using their 1-m and 6-m telescopes. Additionally, the campaign made use of the 2.1-m telescope of Guillermo Haro Astrophysical Observatory (Mexico), 2.1-m telescope of the Observatorio Astronomico Nacional at San Pedro Martir (Mexico), and the 3.5-m and 2.2-m telescopes of Calar Alto Observatory (Spain). 
The spectral data have been collected for several decades (since 1987 until 2015) and have been reduced and calibrated using the same procedures and methods, thus producing a homogeneous set of data. The information about the observations and the description of all procedures for each object are given in \cite{Sh01, Sh04, Sh08, Sh10a, Sh12, Sh13, Sh16, Sh17}. Here we outline some of the important line parameter measurements.  In order to measure the broad line fluxes and full width half maximum (FWHM), we needed to subtract  the underlying continuum emission, and the narrow and other near-by satellite lines, for which we performed the multi-Gaussian fittings \citep{Po04}. In case of H$\beta$ line region, it is important to carefully model the Fe II multiplet emission, and this was done using a template\footnote{For Fe II modeling and subtraction see the page of the Serbian Virtual Observatory, http://servo.aob.rs.} described in details in \cite{Ko10} and \cite{Sh12}.

In Table 1 we list the spectral characteristics of the sample, together with the basic information { (see caption of Table 1)}. The $R_{\rm BLR}$ comes from the time-lags calculated using the  Z-transformed Discrete Correlation Function \citep{Al97}, which was applied on { continuum and line flux light-curves. Depending on the sampling of light curves we used either observed or simulated light-curves in order to get the most reliable result (for details see references in column 9, Table 1)}. The listed masses are taken from the monitoring results (see column 8 in Table 1), { and for details on how the mass was calculated see given references (column 9 in Table 1). In case of NGC 5548 and NGC 4151 the mass was not derived in the original monitoring campaign, so we calculated in this work their SMBH masses, using the equation} $M_{\rm BH}=f \Delta R_{\rm BLR} V^2_{\rm FWHM}/G$, where $\Delta V_{\rm FWHM}$ is the orbital velocity estimated from the width of the variable part of H$\beta$ (taken from references given in Table 1), and $f$ is the virial factor taken to be 5.5 \citep{On04}. The variability parameter $F_{\rm var}$ gives the level of flux variability and is calculated using equation from  \cite{OB98}. For the luminosity at 5100 \AA \, we calculated the  luminosity distances with the online calculator of  \cite{Wr06}, taking the same cosmological parameters as in \cite{Il15}: $\Omega_{\rm } = 0.286$, $\Omega_{\Lambda} = 0.714$, and $\Omega_k = 0$, and a Hubble constant, $H_{\rm 0} = 69.6$ km/s Mpc$^{-1}$.

\begin{table*}[t!]
\caption{Spectral characteristics of the sample. The columns are: (1) object name and monitoring period, (2) redshift, (3) AGN type, (4) shape of line profile and FWHM of the mean H$\beta$, (5) BLR radius $R_{\rm BLR}=c\tau_{\rm BLR}$ of H$\beta$, (6) variability parameter $F_{\rm var}$ of H$\beta$, (7) mean continuum luminosity at 5100\AA, (8) the mass of the SMBH $M_{\rm BH}$, and (9) references.}
\label{tab1}
\vskip 2mm
\centering
\resizebox{18cm}{!}{%
\begin{tabular}{ccccccccc}
\hline
 object  &  z &  AGN   & Line Profile Shape& c$\tau_{\rm BLR}$   &  F$_{\rm var}$& $\lambda$L$_\lambda$(5100)&  
 $M_{\rm BH}$ & References \\
period [years] &    &  type  & FWHM  [km/s]   & H${\beta}$ [ld] &  H${\beta}$ & [$10^{44}$ erg/s] & [$M_\odot$]  &  \\
(1) & (2) & (3) & (4) &(5) &(6) &(7) & (8) &(9) \\
\hline
\hline          
NGC 5548  & 0.0172 & Sy 1.0--1.8& strong shoulders & 49$^{+19}_{-8}$  & 0.33& 0.40$\pm$0.12 & 2.1 $\times 10^9$& \cite{Sh04} \\
 (1996-2002) &        &                &   6300  &   &      &  & & \cite{Kov14} \\
\hline
NGC 4151 & 0.0033 & Sy 1.5--1.8& absorption,bumps  & 5$^{+28}_{-5}$   & 0.42 & 0.05$\pm$0.03 & $1.6 \times 10^8$& \cite{Sh08, Sh10b}  \\
(1996-2006) &        &               &   6110    &  &  &  & & \cite{Ra17}\\
\hline
3C390.3 & 0.0561 & radio loud & double-peaked & 96$^{+28}_{-47}$ & 0.38 & 0.90$\pm$0.42 & $2.1 \times 10^9 $ & \cite{Sh01, Sh10a} \\
(1995-2007) &      &                 &   11900   &   & &  &  & \cite{Jo10, Po11}\\
\hline
Ark 564 & 0.0247 & NLSy1 &  strong FeII & 4$^{+27}_{-4}$  & 0.07 &  0.36$\pm$0.04 & $1.0 \times 10^6$&  \cite{Sh12}\\
(1999-2010)   &        &             &  960    &    &  &&  & \\
\hline
Arp 102B & 0.0242 &LINER &  double-peaked & 15$^{+20}_{-15}$ & 0.21 &0.11$\pm$0.01 & $1.1 \times 10^8$& \cite{Sh13}\\
(1987-2013) &        &              & 15900   &    &   &  &   &  \cite{Po14}     \\
\hline
NGC 7469 & 0.0163   &Sy 1.0 &  narrower & 21$^{+7}_{-0}$ & 0.23 &0.52$\pm$0.0.08  & $1.1 \times 10^7$ & \cite{Sh17}\\
(1996-2015) &        &              &  2000   &    &   &  &   &      \\
\hline
E1821+643 & 0.297   &quasar &  red-asymmetry & 118$^{+0.1}_{-0.0}$ & 0.07  &104.4$\pm$19.9 & $2.6 \times 10^9$ & \cite{Sh16}\\
(1990-2014) &        &              &  5610  &    &   &  &   &  \cite{Kov17}   \\
\hline
\end{tabular}
}
\end{table*}

\begin{figure}[h!]
\begin{center}
\includegraphics[width=12cm]{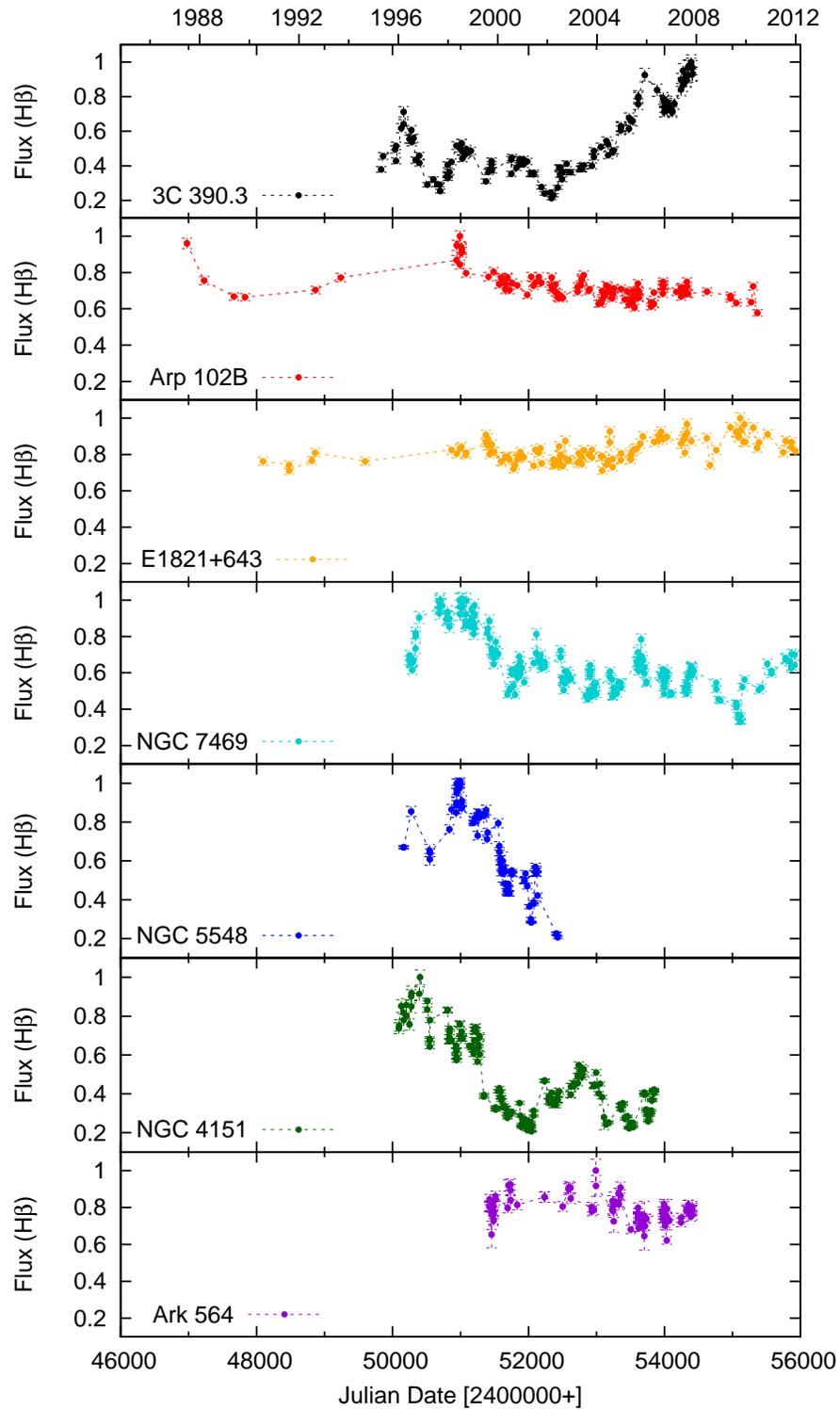}
\end{center}
\caption{Light curves of the H$\beta$ line flux of all object in the sample (object name given in the bottom-left), normalized to the maximal flux for better comparison.}\label{fig:1}
\end{figure}

\begin{figure}[h!]
\begin{center}
\includegraphics[width=10cm]{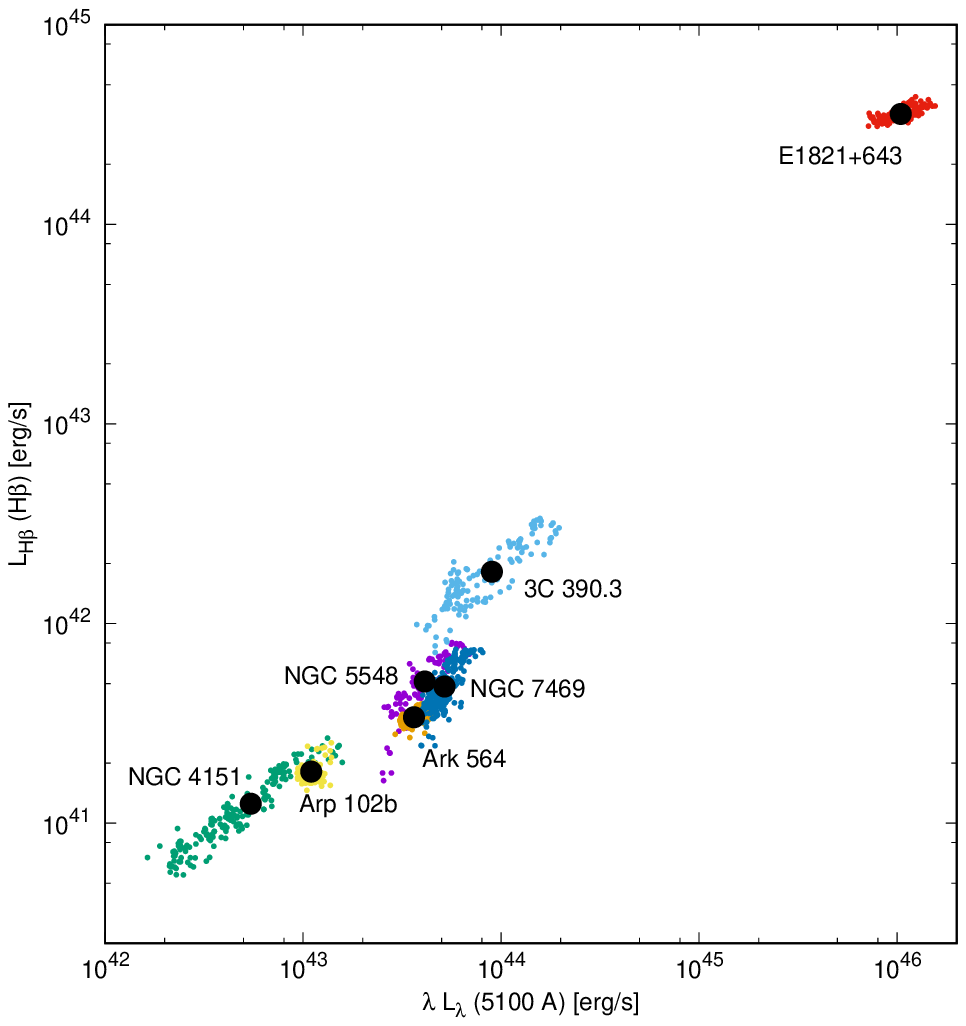}
\end{center}
\caption{H$\beta$ line luminosity for all 7 objects  versus the continuum luminosity at 5100 \AA\, (dots). The mean value is also shown ({ filled} circle) with the object name denoted next to it. }\label{fig:2}
\end{figure}

\section{Results}

In this paper  we present a sample of 7 type 1 AGN with different broad line
profiles (Table 1, columns 3-4) and different variability rate (Table 1, column 6).
{ Figure \ref{fig:1} { shows} the H$\beta$ line flux light curves for all 7 objects, where the line flux and the errors-bars are shown normalized to the maximal flux for comparison. 
In all 7 objects, for which we have been collecting spectra for more than a decade and in some cases two decades, the line and continuum flux are varying. 
From Table 1 and Figure \ref{fig:1}, we can see that:}
NGC 4151 and NGC 5548 are strongly variable Seyfert 1.5, NGC 7469 is also a Seyfert 1 
but with lower variability in line fluxes and profiles, Arp 102B and 3C 390.3 have broad double-peaked lines, Ark 564 is a NLSy1, and E1821+643 is a quasar with strong red-asymmetry in line profiles, but almost no variability. 

One important prerequisite for the estimates of the { radius of the BLR} using the reverberation monitoring data, is the fact that the BLR is photoionized by the central continuum source \citep[see e.g.][]{PH06}. The simplest  way to test this is by plotting the line luminosity as a function of the continuum luminosity. In Figure \ref{fig:2} we thus show the H$\beta$ line luminosity for all 7 objects (shown with dots) versus the continuum luminosity at 5100 \AA. Moreover, the mean value is also shown ({ filled} circle) with the object name denoted next to it. { The important finding is the fact that the luminosity of the H$\beta$ line is following the same trend with the continuum luminosity for all objects, from the low-luminosity Seyfert up to the high-luminosity quasar (Figure \ref{fig:2}). }

\begin{figure}[h!]
\begin{center}
\includegraphics[width=14cm]{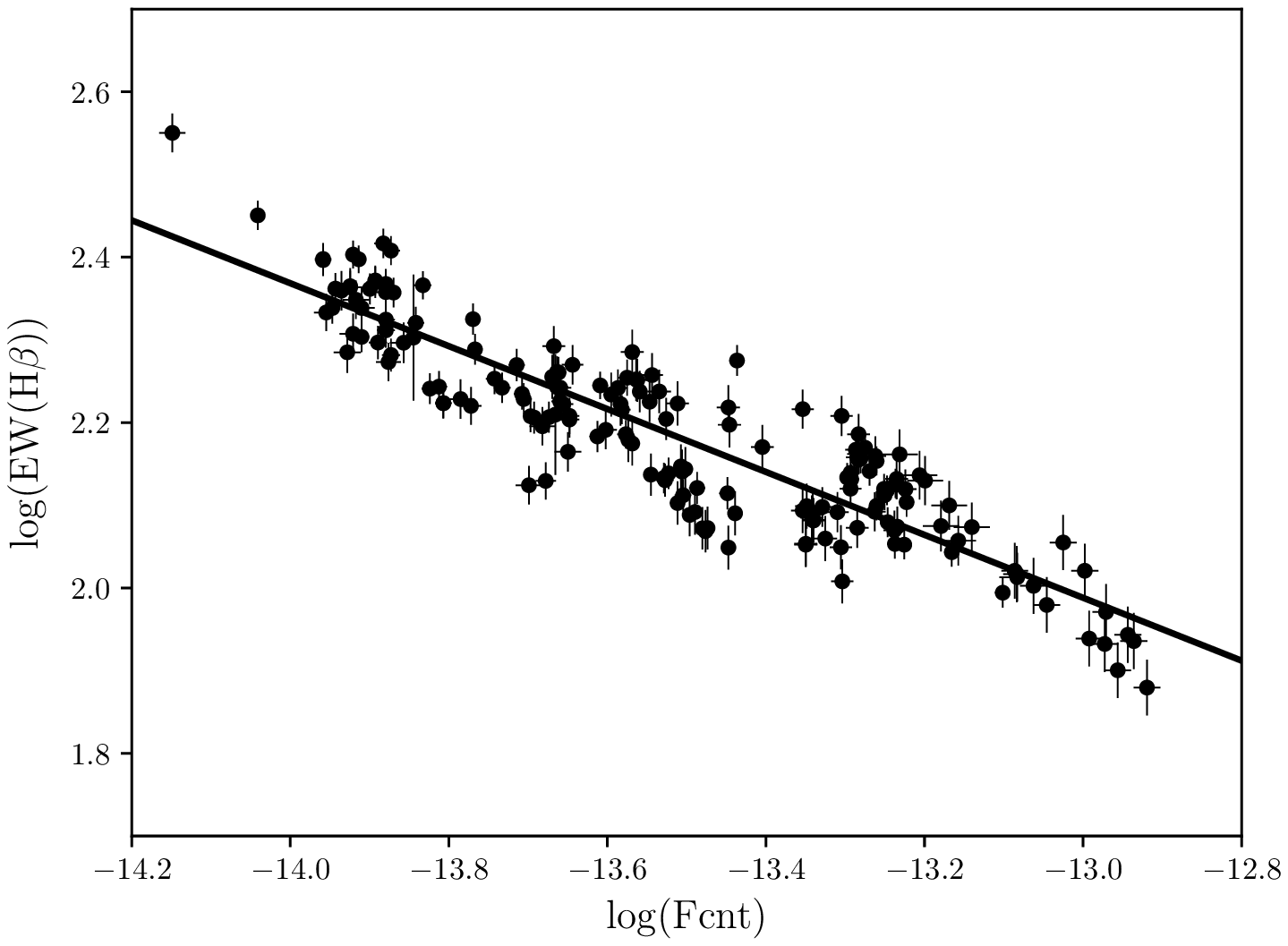}
\end{center}
\caption{ The intrinsic Baldwin effect, i.e. the H$\beta$ line equivalent width vs. the continuum flux at 5100 \AA, in case of Seyfert 1 galaxy NGC 4151.}\label{fig:3}
\end{figure}

On the other hand, it was shown that the intrinsic Baldwin effect is detected in the Balmer lines { (H$\alpha$ and H$\beta$)} in 6 objects from this sample \cite[see the results and discussion in][]{Ra17}. Here we show as an example the presence of the intrinsic Baldwin effect, { an anti-correlation between the H$\beta$ line equivalent width and the continuum flux\footnote{since the intrinsic Baldwin effect is detected in a single object, we can use a flux instead of luminosity} measured at 5100 \AA)} in case of H$\beta$ of NGC 4151 (Figure  \ref{fig:3}), for which this effect is of high significance \citep[see][]{Ra17}.

\begin{figure}[h!]
\begin{center}
\includegraphics[width=14cm]{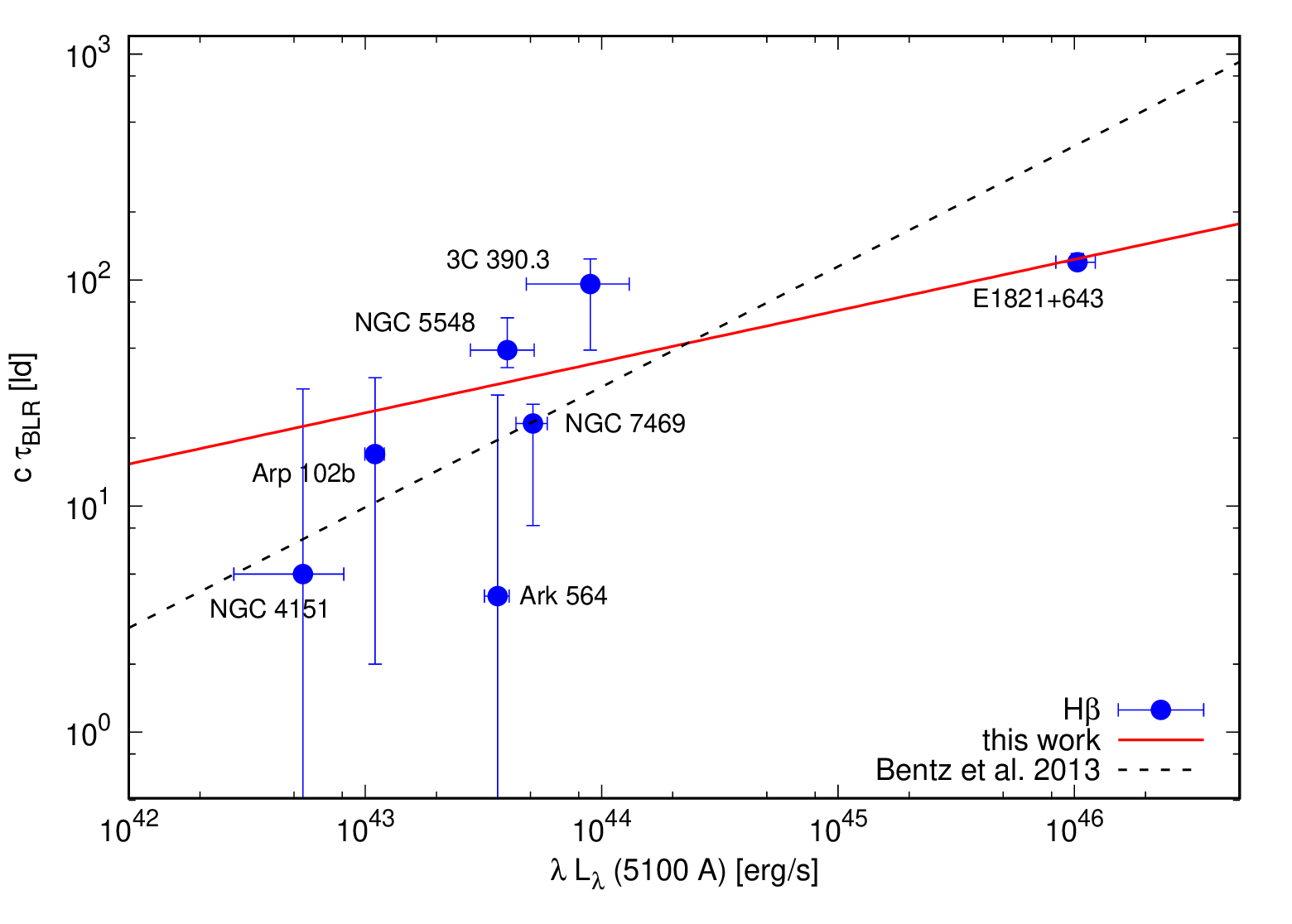}
\end{center}
\caption{{ The radius of the broad H$\beta$ line emitting-region} versus the continuum luminosity at 5100 \AA. The dashed line gives the BLR radius-luminosity scaling relation from \cite{Be13} with $R_{\rm BLR}\sim L^{0.533}$, whereas the solid line is the fit through seven objects from this study, with $R_{\rm BLR}\sim L^{0.226}$. { We note that the fit is strongly influenced from the one point, the luminous quasar E1821+643, and without that point the relation is consistent with the \cite{Be13} result.}}\label{fig:4}
\end{figure}

Figure \ref{fig:4} shows the radius of the broad H$\beta$ line emitting-region as a function of the continuum luminosity at 5100 \AA\, for our sample (Table 1, columns 5 and 7, respectively). Data are fitted with the scaling relation from \cite{Be06} in the form of $\log R_{\rm BLR} = K + \alpha \log [\lambda L_\lambda(5100)/10^{44}]$, where $\alpha$ is the slope of the BLR radius-luminosity relationship and $K$ is the scaling factor. The dashed line is plotted with the coefficients $\alpha=0.533$ and $K=1.527$, given by \cite{Be13}, and the solid line is a simple linear best-fitting of the above equation (not considering the error-bars) through all 7 objects, obtaining the fitting parameters of $\alpha=0.226$ and $K=1.639$. { The linear regression fit was done using a nonlinear least-squares Marquardt-Levenberg algorithm available in gnuplot.}

\section{Discussion and conclusions}

Here we have analyzed and compared the optical spectra for seven type 1 AGN which have different variability properties and characteristics of the broad emission line profiles. This remarkable set of spectral data has been collected for several decades (Table 1, column 1) within the SAO RAS long-term monitoring campaign. Following the comparative study given in \cite{Il15}, we extend here our sample to 7 objects and outline the most important results. We repeat again that all observational data sets were uniformly processed, and in that way could be compared.

If we compare the normalized light-curves shown in Figure \ref{fig:1}, the strongest variable objects are NGC 5548, NGC 4151, and 3C 390.3 (see also in Table 1 the variability parameter F$_{\rm var}$ which is $\sim$40\%), followed by Arp 102B and NGC 7469 (F$_{\rm var}\sim$20\%). The least variable objects are the NLSy1 Ark 564 and the highly luminous quasar E1821+643, for which the line and continuum flux 
are almost constant for decades (for the H$\beta$ line, F$_{\rm var}\sim$7\%). The case of the quasar E1821+643 is specially puzzling since its broad emission lines show a peak shift (of the order $\sim$1000 km/s) and a strong asymmetry in the red wing, extended up to $\sim$15,000 km/s \citep[see][and their Figure 15]{Sh16}. This object was suggested to be a candidate for a binary SMBH or a recoiling black hole after the collision of two SMBH \citep{Ro10}, however \cite{Sh16} found it difficult to explain the unchanging line-profiles on such a long time-scales with the binary SMBH hypothesis, suggesting that the broad-line emitting region may be connected with some sort of gas-rich clouds that are orbiting around the recoiling SMBH.

Interestingly, the rate of variability shows no trend with the SMBH mass (Table 1, column 8), i.e. the low variability is detected in case of AGN with small and large SMBH mass (see Table 1). { This may
be a result of the small number of objects in our sample, since it is well known that Seyfert galaxies, which are low luminosity type 1 AGNs have lower SMBH mass and stronger variability than quasars, which are high luminosity type 1 AGNs with much higher SMBH mass and much lower variability.}

{ The luminosity of H$\beta$ is following the same trend with the continuum luminosity for all objects, regardless of the luminosity.} This is in support of the prediction that the photoionization by the central continuum source is mainly responsible for the heating and line production, which is one of the assumptions needed when calculating the BLR radius from reverberation mapping. However, if considering a single object, the correlation between the line and continuum emission can be violated, which is seen in the case of NGC 4151 and Ark 564 \cite[see for details][respectively]{Sh08, Sh12}. Another correlation that is present and significant is the intrinsic Baldwin effect in the H$\beta$ line. The physical origin of this effect is still unknown \cite[see][for discussion]{Ra17}, but some lines of evidence suggest the possibility that an additional optical continuum component, not related to the ionizing continuum, can be originated either in the BLR or in nuclear outflows, thus influencing the slope of the H$\beta$ - continuum flux relation.

However, what is intriguing is the fact that all presented objects have different line profiles (Table 1, column 4), indicating that the geometry of the broad line-emitting region is probably different. For example, in case of 3C 390.3 it is clear that the BLR is following the accretion disk geometry \citep{Po11}, whereas in case of the other double-peak line emitter Arp 102B this is maybe not the case \citep{Po14}. Also, the Seyfert galaxies NGC 4151 and NGC 5548 could host a binary SMBH with two separate BLR \citep{Bo12, Bo16, Li16}, which puts more uncertainty  on their mass estimates from reverberation mapping. Finally, the quasar E1821+643 is much below the theoretical radius-luminosity relation predicted by the photoionization theory ($R_{\rm BLR} \propto L^{0.5}$), and very low-variable AGN in the optical lines and continuum, therefore this SMBH mass estimate is also more uncertain. The peculiarity of the BLR of this quasar is probably the reason why the slope of the radius-luminosity scaling relation is much smaller than \cite{Be06} relation. It is worth noticing that in Figure \ref{fig:4}, if we discard Ark 564 (with very narrow lines that are not
typical for type 1 AGNs and with a very large error-bars in estimated time-delays) and
E1821+643 (a quasar with an outstanding luminosity), the remaining
objects will apparently closely follow the dependence of \cite{Be13}.

We summarize here the main conclusions of this study:
\begin{itemize}
\item[(i)] the rate of variability is not connected with the geometry of the 
BLR and mass of the SMBH (e.g. Arp 102B and 3C390.3), { however in case of the SMBH mass this
may be a result of the small number of objects in our sample};
\item[(ii)] the luminosity of H$\beta$ is correlated with the continuum luminosity for all AGN in our sample,
and the intrinsic Baldwin effect is present;
\item[(iii)] { the photoionization is in general the main line-production mechanism}, however in some AGN (e.g. NGC 4151) there are some additional mechanism which contribute to the 
line (continuum) emission.
\end{itemize}
Finally we can conclude that the long-term spectroscopic monitoring campaigns are very important for the investigation of the BLR structure and the SMBH mass estimates.

\section{Funding}

RFBR (grants N97-02-17625 N00-02-16272, N03-02-17123, 06-02-16843, N09-02-01136, 12-02-00857a, 12-02-01237a, N15-02-02101); Ministry of Education, Science, Technology and Development, Republic of Serbia (project 176001 "Astrophysical Spectroscopy of Extragalactic Objects"); INTAS (grant N96-0328); DFG grants Ko857/32-2 and Ko857/33-1; Ministry of Science and Technology of R. Srpska (project "Investigation of supermassive binary black holes in the optical and X-ray spectra").


\end{document}